\documentclass[aps,prb,twocolumn,showpacs,superscriptaddress]{revtex4-1}
\usepackage{graphicx}
\usepackage{amsfonts,amssymb,amsmath}
\usepackage{bm}

\usepackage{color}
\newcommand\COMMENTED[1] {}

\begin{document}

\title{Electron-phonon coupling and exchange-correlation effects in superconducting
${\rm H_3S}$ under high pressure}

\date{\today}
 
\author{Matej Komelj}
\email{matej.komelj@ijs.si\\Visiting scientist at the College of William and 
Mary}
\affiliation{Jo\v zef Stefan Institute, Jamova 39, SI-1000 Ljubljana,
  Slovenia}
\author{Henry Krakauer}
\affiliation{Department of Physics, College of William and Mary, 
Williamsburg, VA 23187-8795}
\begin{abstract}

We investigate the ${\rm H_3S}$ phase of sulphur hydride under high pressure $\simeq$ 200 GPa
by means of {\it ab-initio} calculations within the framework of the 
density-functional theory (DFT) with the PBE0 hybrid exchange-correlation ($E_{\rm xc}$)
approximation.
The choice of $E_{\rm xc}$ has the largest effect on the calculated 
electron-phonon coupling (EPC) matrix elements; the high pressure equation of state and phonon
frequencies are only slightly modified. Mode-dependent EPC correction factors are determined from PBE0
using a frozen-phonon supercell approach, while standard density-functional perturbation theory is used
to determine the EPC with
PBE  generalized-gradient approximation $E_{\rm xc}$. Our principle finding is that 
the calculated PBE0 $T_c$ is enhanced by 25\% compared to PBE. 
This is similar in magnitude, but in opposite direction, to the proposed suppression of $T_c$ by anharmonic effects 
[Errea {\em et al., Phys. Rev. Lett.} {\bf 114}, 157004 (2015)]. Our calculations demonstrate the importance of 
considering exchange-correlation approximations for calculations of superconducting properties for this class of materials.

\end{abstract}

\maketitle

The quest for the holy grail of
high-pressure physics, metallic hydrogen, has continued to attract the interest of
experimentalists 
and
theorists 
since Ashcroft 
proposed that the new phase should 
exhibit superconductivity with $T_c\sim 270\>{\rm K}$.\cite{PhysRevLett.21.1748}
A recent focus has been on the hydrides where reduced metallization pressures are 
expected.\cite{PhysRevLett.92.187002}
Sulphur hydrides have attracted a great deal of attention due to theoretical 
predictions\cite{PhysRevLett.92.187002,Li:2014,Duan:2014,PhysRevB.91.060511,PhysRevLett.114.157004,PhysRevB.91.224513} 
of high $T_c\sim 200\>{\rm K}$ under high pressure. 
This was supported by two experimental 
reports:\cite{Drozdov:2014,Drozdov:2015}
superconductivity was first attributed to ${\rm H_2S}$  in Ref.~\onlinecite{Drozdov:2014}, 
but Drozdov {\it et al.}\cite{Drozdov:2015} re-analyzed their measurements, 
which lead to $T_c\approx 203\>{\rm K}$, 
now ascribed 
to the ${\rm H_3S}$ phase. 
Theoretically, Li {\it et al.}\cite{Li:2014}  proposed a metallic phase of hydrogen 
sulfide ${\rm H_2S}$, potentially superconducting with a maximum 
$T_c\sim 80\>{\rm K}$ when subjected to a pressure of 160 GPa.
The transition temperature was estimated by applying the Allen-Dynes 
modified McMillan equation\cite{PhysRevB.12.905} and assuming 
electron-phonon coupling as the source of superconductivity. 
Duan {\it et al.}\cite{Duan:2014} performed a systematic investigation of the 
pressure-dependent ${\rm (H_2S)_2H_2}$ phase diagram and concluded
that the cubic $Im\bar{3}m$ ${\rm H_3S}$ structure  
was the most stable phase at pressures above 180 GPa, with  $T_c\sim 200\>{\rm K}$ at  200 GPa.
Papaconstantopoulos {\it et al.}\cite{PhysRevB.91.184511} applied the Gaspari-Gyorffy 
theory\cite{PhysRevLett.28.801} and argued that  superconductivity in ${\rm H_3S}$
arose mostly from the coupling between the electrons
and the H vibrations, whereas the role of the sulphur is to stabilize the 
hydride at high pressures via hybridization.  A similar hypothesis was put
forward by Bernstein {\it et al.}\cite{PhysRevB.91.060511} who proposed that 
the transport mechanism in the high-pressure ${\rm H_3S}$ was the same
as in ${\rm MgB_2}$. A substantially higher $T_c$ in the hydride could be 
explained with
the considerably smaller atomic masses of the constituents. 
It has also been argued 
that the features of the calculated 
electron-phonon spectrum in ${\rm H_3S}$ are near optimum for a high $T_c$.\cite{PhysRevB.91.220507}  
Errea {\it et al.} \cite{PhysRevLett.114.157004} claimed, however, that the harmonic
approximation, used in the previous calculations, overestimated
the electron-phonon coupling and consequently the predicted $T_c$. 
Other related candidate materials, such as ${\rm SeH_3}$, have also been
investigated.\cite{Flores-Livas:2015}
By contrast, Hirsch and Marsiglio\cite{Hirsch201545} have argued against the conventional electron-phonon mechanism
and in favor of an electron-hole mediated mechanism.

\par
The 
previous theoretical 
investigations\cite{Li:2014,Duan:2014,PhysRevB.91.224513,PhysRevB.91.060511,PhysRevLett.114.157004}
were based on {\it ab-initio} calculations that used the semilocal 
PBE\cite{PhysRevLett.77.3865} 
generalized-gradient approximation (GGA). 
Standard 
local or semilocal DFT approaches sometimes fail
to predict the correct $T_c$ in conventional electron-phonon superconductors because 
of an insufficient treatment of the exchange-correlation 
effects.\cite{PhysRevB.78.081406,PhysRevB.81.073106,PhysRevX.3.021011}
In this paper we address the effect of the choice of DFT 
exchange-correlation functional on the predicted EPC strength.
Hybrid DFT (HDFT) methods 
often improve the predicted properties
of materials with hydrogen bonds.\cite{PhysRevB.58.7701,1742-6596-377-1-012093}
In HDFT a fraction of the
Hartree-Fock exact-exchange term is added
to $E_{xc}$.
\cite{Becke:1993} As a result, HDFT calculations are typically
at least an order of magnitude computationally more demanding than 
local or semilocal $E_{\rm xc}$.
To calculate  the effective Kohn-Sham potential for HDFT,
it is necessary to evaluate 
$N_\mathbf{k}\times N_{occ}$ 
six-dimensional spatial integrals, 
where $N_\mathbf{k}$ and $N_{occ}$ are the number of $\mathbf{k}$-points and
occupied states, respectively. By contrast, only three-dimensional integrals are required
for local or semilocal $E_{\rm xc}$. For this reason, the extension and application of DFT perturbation theory (DFTPT)
to calculate electron-phonon coupling coefficients\cite{PhysRevB.72.024545} becomes very challenging in HDFT.
Instead, we adapted the approximation scheme introduced by 
Lazzeri {\em et al.}~\cite{PhysRevB.78.081406} and 
Yin {\em et al.}~\cite{PhysRevX.3.021011} for GW and HDFT, as described below.

 \par
\begin{figure}
\includegraphics[width=.5\textwidth]{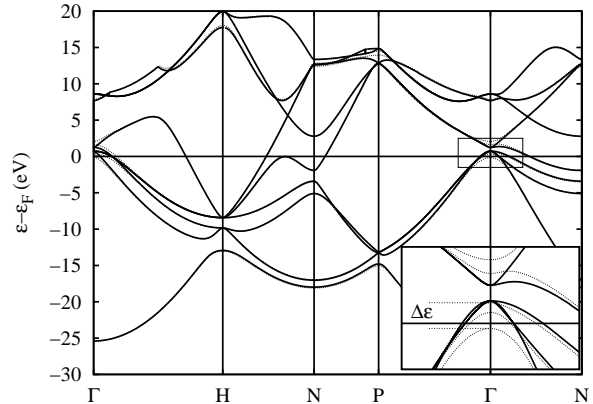}
\caption{Band splitting due the presence of a phonon.
PBE band structures of $Im\bar{3}m$ ${\rm H_3S}$ are shown: solid lines
correspond to the undistorted crystal,
and dashed lines are for the atoms
displaced according to the phonon mode $\nu=6$ at $\mathbf{q}=0$.
The inset presents the details of the band splitting close to the Fermi
energy near the $\Gamma$ point due to the lifted degeneracy caused by the presence
of the phonon.}
\label{fig:delta-e}
\end{figure}

The mode-dependent EPC strengths $\lambda_{\mathbf{q}\nu}$ are given by
\onecolumngrid
\begin{equation}
\lambda_{\mathbf{q}\nu}=\sum_I
 {1\over M_I {\cal N}(\varepsilon_F)\omega_{\mathbf{q}\nu}^2}
\sum_{ij}\int {d^3 k\over\Omega_{\rm BZ}}
\left|\left<\psi_{\mathbf{k},i}\right|
{dV_{\{\mathbf{R}\}}\over d\mathbf{R}_I}\cdot\mathbf{U}_{\mathbf{q}\nu}^I
\left|\psi_{\mathbf{k}+\mathbf{q},j}\right>\right|^2
\delta(\varepsilon_{\mathbf{q},i}-\varepsilon_{F})
\delta(\varepsilon_{\mathbf{k}+\mathbf{q},j}-
\varepsilon_F),
\label{eq:EPC}
\end{equation}
\twocolumngrid
\noindent 
where the phonon frequencies $\omega_{\mathbf{q}\nu}$  and eigenvectors $\mathbf{U}_{\mathbf{q}\nu}^I$ are
obtained by diagonalizing the phonon dynamical matrix; $\varepsilon_{\mathbf{k},i}$, 
$\psi_{\mathbf{k},i}$  are the single-electron energies and eigenfunctions,
${\cal N}(\varepsilon_F)$ is the density of states at the Fermi energy
$\varepsilon_F$, and $M_I$ is the mass of the $I$th atom. The EPC matrix element
$\left<\psi_{\mathbf{k},i}\right|
{dV_{\{\mathbf{R}\}}\over d\mathbf{R}_I}\cdot\mathbf{U}_{\mathbf{q}\nu}^I
\left|\psi_{\mathbf{k}+\mathbf{q},j}\right>$ depends on the change of the self-consistent Kohn-Sham potential 
with respect to the phonon displacement. All of the quantities in Eq.~(\ref{eq:EPC}) can be calculated using 
DFTPT\cite{PhysRevLett.58.1861,PhysRevB.55.10355,PhysRevB.72.024545}
for  local or semi-local $E_{\rm xc}$, and this functionality is  
 available in standard codes such a Quantum 
Espresso.\cite{Giannozzi:2009} Unfortunately, a combination of
HDFT and DFPT would be computationally too demanding and has not yet been 
implemented for solids, to our knowledge. 
Instead, we use an alternative approach.\cite{PhysRevB.78.081406,PhysRevX.3.021011}  
Supercells and the frozen-phonon method are used to obtain the phonon dynamical matrix and the 
resulting pairs of $\omega_{\mathbf{q}\nu}$, $\mathbf{U}_{\mathbf{q}\nu}^I$
on a $\mathbf{q}$-grid 
commensurate with the supercell.\cite{phonopy}
Small but finite phonon displacements can then be introduced to obtain
phonon-induced Kohn Sham potentials, which can be expressed to first order as:
$V_{\{\mathbf{R}\}}+(dV_{\{\mathbf{R}\}}/d\mathbf{R}_I)\cdot
\mathbf{U}_{\mathbf{q}\nu}^I$
The change 
$\Delta\varepsilon$ of a single-electron energy is approximated as:
\begin{equation}
\Delta\varepsilon\approx\left<{dV_{\{\mathbf{R}\}}\over d\mathbf{R}_I}\cdot
\mathbf{U}_{\mathbf{q}\nu}^I\right>. 
\label{eq:delta-e}
\end{equation}
The difference $\Delta\varepsilon$ describes band splitting due to the lifted degeneracy 
in the presence of the phonon.
For the application of the approximation (\ref{eq:delta-e}) for the 
calculation of $\lambda_{\mathbf{q}\nu}$ only
the splittings 
nearest to the Fermi level are relevant. An example is presented in Fig.~\ref{fig:delta-e}
for $Im\bar{3}m$ ${\rm H_3S}$ in the presence of phonon mode $\nu=6$ at $\mathbf{q}=0$,
calculated with PBE $E_{xc}$.
\par
The band splittings are 
determined for both
DFT and HDFT, 
$\Delta\varepsilon^{\rm DFT}_{\mathbf{q}\nu}$ and 
$\Delta\varepsilon^{\rm HDFT}_{\mathbf{q}\nu}$, respectively.  
Using Eq.~(\ref{eq:delta-e}) in Eq.~(\ref{eq:EPC})
for DFT and for HDFT yields the following approximation
to estimate the HDFT electron-phonon coupling coefficients:
\begin{equation}
\lambda_{\mathbf{q}\nu}^{\rm HDFT}\approx\lambda_{\mathbf{q}\nu}^{\rm DFT}
f_{\mathbf{q}\nu},
\end{equation}
where the correction factor $f_{\mathbf{q}\nu}$ is given by:
\begin{equation}
f_{\mathbf{q}\nu}=
{{\cal N}^{\rm DFT}\left(\varepsilon_F^{\rm DFT}\right)\over 
{\cal N}^{\rm HDFT}\left(\varepsilon_F^{\rm HDFT}\right)}
\left(
{\omega_{\mathbf{q}\nu}^{\rm DFT}\over\omega_{\mathbf{q}\nu}^{\rm HDFT}}
\right)^2
\left(
{\Delta\varepsilon^{\rm HDFT}_{\mathbf{q}\nu}
\over\Delta\varepsilon^{\rm DFT}_{\mathbf{q}\nu}}
\right)^2.
\end{equation}
This procedure 
could be repeated, in principle,
to determine  $\lambda_{\mathbf{q}\nu}^{\rm HDFT}$
for each phonon mode throughout the Brillouin zone (BZ), using appropriate
$\mathbf{q}$-commensurate supercells. 
We note, however, that
the Allen-Dynes 
equation
for 
the superconducting transition temperature\cite{PhysRevB.12.905} 
\begin{equation}
T_C={\omega_{\rm log}\over 1.2}\exp{\left(-1.04(1+\lambda)\over
\lambda-\mu^\ast(1+0.62\lambda)\right)}
\label{eq:Allen-Dynes}
\end{equation}
depends on the total electron-phonon coupling
coefficient $\lambda$, which is given 
by an integral over the BZ 
and sum over phonon branches:
\mbox{$\lambda=(1/\Omega_{\rm BZ})\sum_\nu\int \lambda_{\mathbf{q}\nu}d^3 q$},
where $\Omega_{\rm BZ}$ is the volume of the BZ; $\mu^\ast$ and $\omega_{\rm log}$ are discussed below. 
Therefore we determine the center of gravity $\mathbf{q}_\nu^\ast$ in the irreducible
wedge of the Brillouin zone (IBZ) with respect to 
the $\lambda_{\mathbf{q}\nu}^{\rm DFT}$ for each DFT phonon branch
$\nu$:
\begin{equation}
\mathbf{q}_\nu^\ast={
\int_{\rm IBZ} {d^3 q}
\, \lambda_{\mathbf{q}\nu}^{\rm DFT} \mathbf{q}
\over
\int_{\rm IBZ} {d^3 q}
\, \lambda_{\mathbf{q}\nu}^{\rm DFT}
}.
\end{equation}
For each branch we calculate the correction factor (4)
$f_{\mathbf{q}^\ast\nu}$ and use this single value for all 
$\mathbf{q}$-points within a given branch $\nu$: 
\begin{equation}
\lambda_{\mathbf{q}\nu}^{\rm HDFT}\approx\lambda_{\mathbf{q}\nu}^{\rm DFT}
f_{\mathbf{q}^\ast\nu}
\label{eq:lambda_HDFT}
\end{equation}

\par
The calculations were carried out by applying the Quantum 
Espresso\cite{Giannozzi:2009} code. The DFT exchange-correlation potential
was PBE, whereas for HDFT we used PBE0.\cite{Perdew:1996,Adamo:1998}
The bare electron-ion interactions were described with the norm-conserving
Goedecker-Hartwigsen-Hutter-Teter\cite{PhysRevB.58.3641,PhysRevB.54.1703}
pseudopotentials.  The planewave and charge-density cut-off parameters
were set to $476\>{\rm eV}$ and $1904\>{\rm eV}$, respectively. 
A Monkhorst-Pack\cite{Monkhorst:1976} $16\times 16\times 16$  
$\mathbf{k}$-point grid 
was used for the BZ integration 
primitive cell calculations.
PBE0 phonon spectra were calculated using the frozen-phonon method as
implemented in the Phonopy\cite{phonopy} code, whereas the DFT 
electron-phonon coefficients were determined using DFPT,
which is a part of the Quantum Espresso package.  The total 
electron-phonon coupling coefficient $\lambda^{\rm DFT}$ was obtained 
by using a  
$4\times 4\times 4$ $\mathbf{q}$-point grid in the BZ.
The force constants for the frozen-phonon method were 
obtained from a $2\times 2\times 2$-supercell calculation.

\par
Fig. 2 presents the calculated total energies as a function of the 
lattice parameter $a$ fitted with  the Birch-Murnaghan equation of 
state.\cite{PhysRev.71.809}  
\begin{figure}
\includegraphics[width=.5\textwidth]{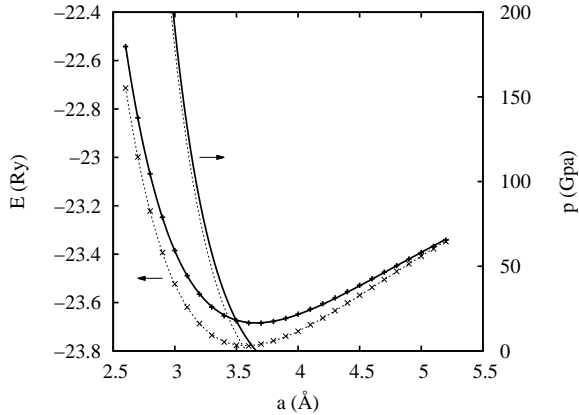}
\caption{The calculated total energy with respect to the lattice 
parameter $a$ for PBE ($+$) and PBE0 ($\times$) fitted with  the 
Birch-Murnaghan equation of state, from which the pressure was calculated. The solid 
lines are for PBE, the dashed lines are for PBE0.}
\end{figure}
The difference between PBE and PBE0 is most pronounced at zero
pressure, reflecting the 
$\simeq 2\%$ difference in equilibrium lattice 
parameters: $3.66\>{\rm\AA}$ and $3.58\>{\rm\AA}$, respectively. 
Near $200\>{\rm GPa}$, however, 
PBE and PBE0 yield very similar lattice parameters, which match
the published\cite{Duan:2014}  value $2.984\>{\rm\AA}$.
We used this value for the subsequent calculations of the phonon-related 
properties. 
\par
The effect of HDFT is larger for phonon frequencies near $200\>{\rm GPa}$, as shown in Fig. 3.
\begin{figure}
\includegraphics[width=.5\textwidth]{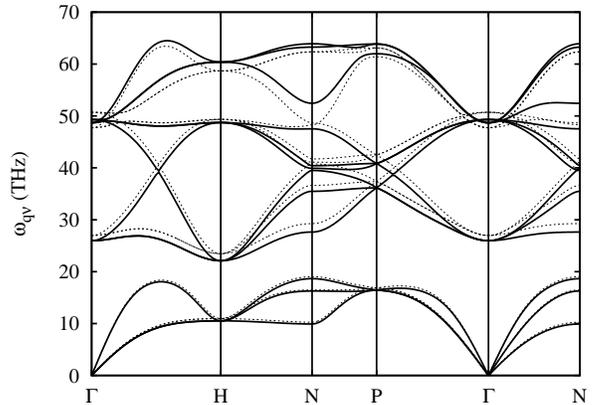}
\caption{Calculated phonon dispersion for the PBE (solid lines) and 
PBE0 (dashed lines) exchange-correlation potentials.}
\end{figure}
The largest differences are for the optic modes, especially near the N point, while
differences for the acoustic modes 
are small.
PBE EPC strengths throughout the Brillouin zone are depicted  
in Fig. 4. 
\begin{figure}
\includegraphics[width=.5\textwidth]{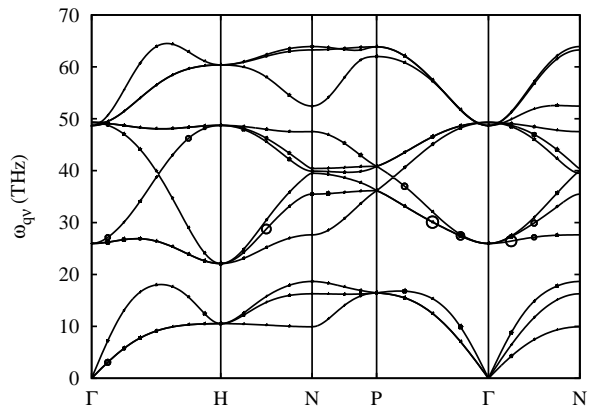}
\caption{
Electron-phonon $\mathbf{q}$-dependent PBE coupling strengths: the size of the circles, superimposed on the
phonon dispersion curves, 
is proportional to the electron-phonon coupling coefficient
$\lambda_{\mathbf{q}\nu}^{\rm DFT}$.}
\end{figure}
In contrast to the case of ${\rm BaBiO_3}$ from 
Ref. \onlinecite{PhysRevX.3.021011}, there are no distinguished points 
with particularly large values of $\lambda_{\mathbf{q}\nu}^{\rm DFT}$. 
This provides some justification for our 
introduction here of the center of gravity point
 $\mathbf{q}^\ast$ in Eq.~(5).
Using this approximation, the number of HDFT $f_{\mathbf{q}\nu}$ evaluations
is reduced from 96 (the number of $\mathbf{q}$-points in the irreducible
BZ times the number of bands) to only 12 (the number of bands) in Eqs. (4) and (6).
For all phonon branches,
the center of gravity point 
is well-approximated by either $\mathbf{q}^\ast=\Gamma$ or \mbox{$\mathbf{q}^\ast=\left(0,0,1/2\right) 2\pi/a$}. 
The band splittings $\Delta\varepsilon$
can therefore be obtained from 
primitive cell or $1\times 1\times 2$-supercell calculations,
respectively. 

A suitable indicator
for the overall influence of the exchange-correlation effects on
the electron-phonon coupling is the Eliashberg spectral function
\mbox{$\alpha^2F(\omega)\propto\sum_\nu\int (1/\Omega_{\rm BZ})
\lambda_{\mathbf{q}\nu}\omega_{\mathbf{q}\nu}\delta(\omega-
\omega_{\mathbf{q}\nu})d^3 q$} which is plotted in Fig. 5 
(using gaussians with width $\sigma=0.5\>{\rm THz}$ to represent the $\delta$-functions).
\begin{figure}
\includegraphics[width=.5\textwidth]{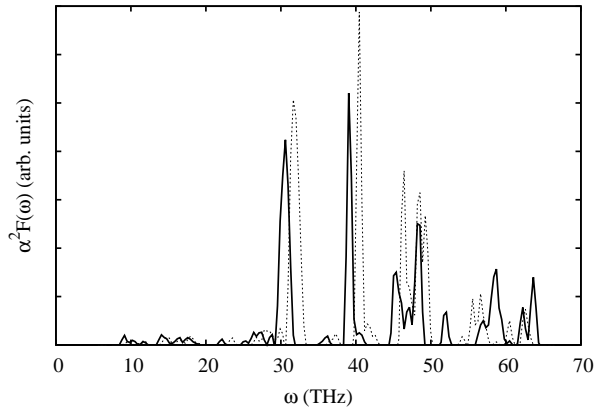}
\caption{The calculated Eliashberg spectral function 
$\alpha^2F(\omega)$, using PBE (solid lines) and
the PBE0 (dashed lines) exchange-correlation potentials.}
\end{figure}
The 
difference in peak positions reflects the
changes in phonon dispersion in Fig.~3. The highest peaks 
are located in the middle of the frequency range, which is in agreement 
with the distribution of the $\lambda_{\mathbf{q}\nu}^{\rm PBE}$ magnitudes 
in Fig. 4. The PBE0 peaks for frequencies up to  $50\>{\rm THz}$ are 
higher than the corresponding PBE peaks, which 
is due to the enhanced phonon-induced band splitting in PBE0 [Eq. (4)]. 
The situation is the opposite at the frequencies
higher than 50 THz, where the PBE peaks are higher. 

The quantitative results are 
summarized in Table I. 
\begin{table}
\vspace{12pt}
\begin{ruledtabular}
\begin{tabular}{lrrr}
&$\lambda$&$\omega_{\log}[K]$&${\rm T}_c[K]$\\
\hline
PBE&1.76&1657&217-201\\
PBE0&2.18&1773&270-253
\end{tabular}
\end{ruledtabular}
\caption{A comparison between the PBE and PBE0 calculated quantities, which 
appear in 
the Allen-Dynes equation (\ref{eq:Allen-Dynes}), and between the resulting $T_c$ for the
retarded Coulomb repulsion $\mu^\ast=0.1-0.13$.}
\end{table}
The superconducting transition temperature $T_c$ in
the Allen-Dynes equation (\ref{eq:Allen-Dynes})
depends not only on the electron-phonon coupling coefficient $\lambda$,
but also on the logarithmic average frequency
$\omega_{\rm log}=(1/\lambda\Omega_{\rm BZ})\sum_\nu\int
\lambda_{\mathbf{q}\nu}
\log(\omega_{\mathbf{q}\nu})d^3 q
$ and the retarded Coulomb repulsion $\mu^\ast$. 
Here, we simply adopt literature values for $\mu^\ast$.\cite{PhysRevB.91.184511} 
From visual inspection of Fig. 3 it is 
not evident whether PBE0 
yields higher or lower phonon frequencies on average than PBE.  
Indeed, the calculated $\omega_{\rm log}$ values differ by only 7\%. 
Thus, the change in phonon frequencies has 
only a small effect 
on the PBE0 enhancement of $\lambda$ by 24\%. 
A comparable enhancement of $\lambda$  was also found for 
graphene/graphite\cite{PhysRevB.78.081406}
and  ${\rm C_{60}}$ molecule.\cite{PhysRevB.81.073106}
As shown in the Table, the calculated PBE0 $T_c$, 
using the Allen-Dynes equation, is higher by about 25\% 
compared to PBE.  
The influence of the exchange-correlation effects on the $\mu^\ast$ is 
complex, and it is not in the focus of the present paper. 
DFT and Hartree-Fock (and thus also HDFT) wave functions differ only slightly,
\cite{doi:10.1021/ja9826892} so the 
double Fermi-surface averaged electron-electron Coulomb interactions will
be similar. Since these determine $\mu$ and $\mu^\ast$ \cite{Allen1982}, 
it is reasonable
to compare predictions of DFT and HDFT using the same $\mu^\ast$. 
On the
basis of the presented results it is
clear that the predicted $T_c$ would be enhanced for any reasonable change in 
$\mu^\ast$ after switching from PBE to PBE0.

The context for our results is the theory of conventional electron-phonon superconductivity as described by DFT.
In H$_3$S and related sulfides, the 
normal-state Fermi liquid quasiparticles have been described
by DFT band structures and wave functions, using PBE $E_{\rm xc}$. 
The DFT approximation is often quite good and is widely used, although 
there are no guarantees regarding the accuracy of the related one-particle 
Kohn-Sham eigenstates.
Phonon excitations are 
commonly
described
in the harmonic approximation, using DFPT or DFT supercell calculations. 
These are ground state properties, 
within the Born-Oppenheimer approximation, so DFT is well
justified. Using DFT for the non-adiabatic EPC is also justified within Migdal's theorem,\cite{Migdal:1958,Allen1982}
and these are treated, as described above. 
The superconducting transition temperature $T_c$ is then usually estimated using the McMillan or Allen-Dynes equations,
although the superconducting anisotropic gap 
equation can be used\cite{PhysRevB.72.024545,2007_SCDFT_aniso-gap,PhysRevB.91.224513,Flores-Livas:2015}  by means of SCDFT.
The SCDFT calculation with harmonic phonons \cite{Flores-Livas:2015} 
yielded $T_c=$180 K for H$_3$S, in agreement with experiment. While SCDFT provides a sound footing for the DFT treatment of superconductivity,  it does 
not in itself improve the electronic band structure and 
phonon properties.
The effects of anharmonicity were studied by Errea {\em et al.}\cite{PhysRevLett.114.157004} 
They found that anharmonicity suppressed $\lambda$ by 30\%; $T_c$
was suppressed by 34 and 56 K, respectively, using the McMillan equation or the isotropic Migdal-Eliashberg equations. 
All of these studies were based on PBE $E_{\rm xc}$.
Here we have shown that the choice of exchange-correlation potential can have effects of similar magnitude on the EPC and $T_c$. 
The present PBE0 calculations modify the electronic quasiparticles in the theory, the phonon excitations, and the EPC. For $T_c$,
we relied on the Allen-Dynes equations with our calculated PBE0 $\lambda$ and phonon frequencies.

\par
In summary, we examined the  influence of exchange-correlation effects 
on the electron-phonon coupling in cubic $Im\bar{3}m$ 
${\rm H_3S}$ under high pressure. We introduced 
the electron-phonon coupling center-of-gravity point $\mathbf{q}^\ast$,
 modifying the approach used in 
Ref. \onlinecite{PhysRevX.3.021011}. 
We calculated a 25\% enhancement of $T_c$ for HDFT compared to PBE predictions.
This is similar in magnitude, but in opposite direction, to the proposed suppression of $T_c$ by anharmonic effects 
in Ref.~\onlinecite{PhysRevLett.114.157004}.
Our results demonstrate the importance of 
considering exchange-correlation approximations for calculations of superconducting properties for this class of materials.

\begin{acknowledgments}
We thank Dimitrios A. Papaconstantopoulos for helpful discussions and Eric J. Walter for consultations
on the computer calculations.
HK acknowledges support from ONR (N000141211042).  We also acknowledge computing support from 
the computational facilities at William and Mary.
\end{acknowledgments}

\bibliography{text.bib}
\end{document}